\documentstyle[preprint,aps]{revtex}
\begin{document}

\tightenlines

\draft

\title{Isotopic Scaling in Nuclear Reactions}

\author{M.~B. Tsang$^{1}$, W.A. Friedman$^{2}$, C.K. Gelbke$^{1}$, W.G. Lynch$^{1}$,
G. Verde$^{1}$, H. Xu$^{1}$}
\author{\small $^{1}$ National Superconducting Cyclotron Laboratory and Department of
Physics and Astronomy, Michigan State University} 
\author{\small $^{2}$ Department of Physics, University of Wisconsin, Madison, WI 53706}


\date{\today}

\maketitle

\begin{abstract}
A three parameter scaling relationship between isotopic distributions for
elements with Z$\leq 8$ has been observed that allows a simple description
of the dependence of such distributions on the overall isospin of the
system. This scaling law (termed iso-scaling) applies for a variety of
reaction mechanisms that are dominated by phase space, including
evaporation, multifragmentation and deeply inelastic scattering. The origins
of this scaling behavior for the various reaction mechanisms are explained.
For multifragmentation processes, the systematics is influenced by the
density dependence of the asymmetry term of the equation of state.
\end{abstract}
\pacs{}

The availability of high intensity radioactive beams facilitates the
exploration of the isospin degree of freedom in nuclear reactions.
Understanding the connection between the entrance channel isospin and the
isotopic distribution of reaction products is important for studying the
charge asymmetry term of the nuclear equation-of-state[1-3], obtaining
information about charge equilibration[4-6], providing stringent tests for
reaction models and optimizing the production of rare isotopes far from
stability. In this letter, we demonstrate that isotopic distributions for
statistical production mechanisms follow scaling laws. We also find
circumstances where the values for the scaling parameters are influenced by
the density dependence of the asymmetry term of the nuclear equation of
state, a quantity that influences many important properties of neutron stars.

The scaling laws in question relate ratios of isotope yields measured in two
different nuclear reactions, 1 and 2, $R_{21}(N,Z)=Y_{2}(N,Z)/Y_{1}(N,Z)$.
In multifragmentation events, such ratios were shown to obey an exponential
dependence on the neutron and proton number of the isotopes characterized by
three parameters $\alpha ,\beta $ and $C$ [7]:

\begin{equation}
R_{21}(N,Z)=C\cdot \mathrm{exp}(\alpha N+\beta Z)  \label{eq:alphabeta}
\end{equation}

\noindent Here we choose the convention that the isospin composition
(neutron to proton ratio) of system 2 is larger than that of system 1. The
systematics described by Eq. \ref{eq:alphabeta} occur naturally within the
grand-canonical ensemble [7-9]. As shown in Ref. [7], the parameters $\alpha 
$ and $\beta $ in that limit are the differences between the neutron and
proton chemical potentials for the two reactions (i.e., $\alpha =\Delta \mu
_{n}/T$ and $\beta =\Delta \mu _{p}/T$ ) and $C$ is an overall normalization
constant.

The accuracy of the iso-scaling described by Eq. \ref{eq:alphabeta} can be
compactly displayed if one plots the scaled isotopic ratio,

\begin{equation}
S(N)=R_{21}(N,Z)\cdot \mathrm{exp}(-\beta Z\;)
\end{equation}
as a function of $N$. For all elements, $S(N)$ must lie along a straight
line on a semi-log plot when Eq. \ref{eq:alphabeta} accurately describes the
experimental data. The data points marked as ''multifragmentation'' in
Figure 1 show values of $S(N)$ extracted from isotope yields with $1\leq
Z\leq 8$ measured for multifragmentation events in central $^{124}Sn+^{124}Sn
$ and $^{112}Sn+^{112}Sn$ collisions at E/A= 50 MeV [7]. Selection of
central events ensures that the average excitation energies and temperatures
in the participant source should be nearly identical  [10]. The observed
iso-scaling is a necessary condition for the applicability of equilibrium
models; such models have described other aspects of these collisions quite
well [11].

Rather surprisingly, iso-scaling is also observed for strongly damped binary
collisions ($^{16}O$ induced reactions on two targets $^{232}Th$ and $%
^{197}Au$) [12] and evaporative compound nuclear decay ($^{4}He+^{116}Sn$
and $^{4}He+^{124}Sn$ collisions)[13], for which Grand-Canonical Ensemble
approaches would appear to have little relevance. Our studies suggest that
iso-scaling is obeyed where 1.) both reactions 1 and 2 are accurately
described by a specific statistical fragment emission mechanism and 2.) both
systems are at nearly the same temperature. Indeed for deeply inelastic
reactions emitted forward of the grazing angles, ($^{22}Ne+^{232}Th$ and $%
^{22}Ne+^{97}Zr$ at $\theta =12{{}^\circ}$ and $E/A=7.9MeV$ [12]) and for
reactions with different temperatures ($^{124}Sn+^{124}Sn$ [7] and $%
^{4}He+^{124}Sn$ [13]), iso-scaling is violated. Conditions 1 and 2 are met
by the three reactions shown in Figure 1. Why iso-scaling is specifically
observed in these cases and what aspects of statistical physics such scaling
probes are examined below.

We first examine the strongly damped collisions, where iso-scaling is
reasonably well respected at low incident energies ($E/A<10MeV$) and at
relatively backward angles [12,14], i.e., when equilibrium is established
between the orbiting projectile and target. In such cases, the isotopic
yields follow the ''$Q_{gg}$-systematics''[12,14], and can be approximated by

\begin{equation}
Y(N,Z)\propto \mathrm{exp}((M_{P}+M_{T}-M_{P}^{\prime }-M_{T}^{\prime })/T)
\label{eq:Qgg}
\end{equation}

\noindent where $M_{P}$ and $M_{T}$ are the initial projectile and target
masses, and $M_{P}^{\prime }$ and $M_{T}^{\prime }$ are the final masses of
the projectile- and target-like fragment. Here, $T$ can be interpreted as
the temperature. Eq. \ref{eq:Qgg} reproduces the systematics shown in Fig.
1. To show why this is so, we have expanded the nuclear binding energy
contributions to the masses in Taylor series in $N$ and $Z$. Expressing
explicitly only the leading order terms that depend on $N$ and $Z$, we
obtain a relatively accurate leading order approximation to Eq. \ref{eq:Qgg}
: 
\begin{equation}
R_{21}\propto \exp [(-\Delta s_{n}\cdot N-\Delta s_{p}\cdot Z)/T].
\label{eq:R21_Qgg_sep}
\end{equation}

\noindent where $\Delta s_{n}$ and $\Delta s_{p}$ are the differences of the
neutron and proton separation energies for the two compound systems. Thus,
the difference in the average separation energies in Eq. \ref{eq:R21_Qgg_sep}
plays a corresponding role to the difference in chemical potentials in the
grand canonical approach, an intriguing result when one considers that $\mu
\approx -s$ in the low temperature limit [15]. The straightforward
dependence of Eq. \ref{eq:R21_Qgg_sep} on temperature suggests that it may
provide information relevant to the temperatures achieved in strongly damped
collisions. The data of Figure 1 imply a temperature of 2.7 MeV, not
inconsistent with values derived from alternative analyses [12,14].

Next we consider the yields from processes involving the formation of a
composite system and its subsequent decay via the evaporation of different
isotopes. Corresponding scaled isotopic ratios for fragments detected at
backward laboratory angles ($\theta =160^{o}$) in $^{4}He+^{116}Sn$ and $%
^{4}He+^{124}Sn$ collisions at $E/A=50MeV$ [13] are shown in Figure 1, next
to the label ''evaporation''.

To explore the factors that govern the relevant evaporation rates, we
utilize the formalism of Friedman and Lynch [16] which invokes statistical
decay rates derived from detailed balance [17]. When the yields are
dominated by emission within a particular window of source-mass or
source-temperature, the relative yields of a fragment with neutron number $N$
and proton number $Z$ are directly related to the instantaneous rates 
\begin{eqnarray}
dn(N,Z)/dt &\propto &T^{2}\cdot \mathrm{exp}(-V_{c}/T+N\cdot
f_{n}^{*}/T+Z\cdot f_{p}^{*}/T  \label{eq:Weiss} \nonumber \\
&&-\left\{ BE(N_{i},Z_{i})-BE(N_{i}-N,Z_{i}-Z)-BE(N,Z)\right\} /T)
\end{eqnarray}
where $V_{c}$ gives the Coulomb barrier, the terms $f_{n}^{*}$ ($f_{p}^{*}$)
represent the excitation contribution to the free energy per neutron
(proton),\ BE is the binding energy and $N_{i}$ and $Z_{i}$ identify the
neutron and proton numbers of the parent nucleus.

Applying Eq. \ref{eq:Weiss} to the calculations of $R_{21}$ for two systems
at the same temperature, we find that the binding energies of the emitted
fragments cancel and the systematics shown in Fig. 1 can be reproduced. To
understand why, we again expand the binding energy of the residue with
neutron number $N_{i}-N$ and proton number $Z_{i}-Z$ to leading order in a
Taylor series to obtain: 
\begin{equation}
R_{21}(N,Z)\propto exp[\{(-\Delta s_{n}+\Delta f_{n}^{*})\cdot N+(-\Delta
s_{p}+\Delta f_{p}^{*}+e\Delta \Phi (Z_{i}-Z))\cdot Z\}/T]
\label{eq:R21_Weiss}
\end{equation}

\noindent where $\Phi (Z)$ is the electrostatic potential at the surface of
a nucleus with neutron and proton number $N$ and $Z.$ Aside from the second
order term from the electrostatic potential, which is small for the decay of
large nuclei, all factors in the exponent are proportional to either $N$ or $%
Z$, consistent with Eq. (1). The corresponding scaling parameters $\alpha $
and $\beta ,$ are functions of the separation energies, the Coulomb
potential and small contributions from the free excitation energies. Using
the functions of $\Delta s_{n}$ and $\Delta f_{n}^{*}$ from Friedman and
Lynch [16], one finds that a fixed temperature of about 3.7 MeV is required
in Eq. \ref{eq:R21_Weiss}, to obtain the experimental value of $\alpha =0.6.$
Running a full evaporation chain, using the procedure of Ref. [16], provides
an average fragment emission temperature of about 3.3 MeV. These temperature
values are comparable to those extracted by other techniques [13,18].

The Expanding Evaporating Source (EES) model [19] provides an alternative
description of multifragmentation. Within the context of that model,
additional insights can be obtained. The EES model utilizes a formula for
the particle emission rates which is formally identical to that of Eq. \ref
{eq:Weiss} but can differ significantly in its predictions because the
residue may expand to sub-saturation density [20]. In this circumstance, the
term enclosed in brackets ''\{\}''containing three binding energies in Eq. 
\ref{eq:Weiss} may vanish or become negative, enhancing the emission rate of
fragments with $3\leq Z\leq 20.$ Detailed examination reveals that $\Delta
f_{n}^{*}$ in Eq. \ref{eq:R21_Weiss} is usually much smaller than $\Delta
s_{n},$ and the volume, surface, and Coulomb contributions to $\Delta s_{n}$
largely cancel, leaving the asymmetry energy term, $Sym(\rho )\cdot
(N-Z)^{2}/A,$ alone as the dominant contribution to $\alpha $. For
simplicity, we assume a power law dependence for $Sym(\rho ),$ i.e. $%
Sym(\rho )=C_{sym}\cdot (\rho /\rho _{0})^{\gamma }$ where $\gamma $ is a
variable and $C_{sym}=23.4$ MeV is the conventional liquid drop model
constant [21].

For illustration, we have performed calculations for the decay of the
composite systems found in $^{124}Sn$ + $^{124}Sn$ and $^{112}Sn$ + $^{112}Sn
$ collisions assuming, for simplicity, initial systems of ($Z_{tot},$ $%
A_{tot})$ of (100, 248$)$ and (100, 224), respectively, initial thermal
excitation energies of $E_{ther}^{*}=9.5MeV,$ and initial collective radial
expansion energies of $E_{Coll}/A=2.5MeV.$ The fragments are emitted from
these systems as they expand from an initial density $\rho /\rho _{0}=1$ to $%
\rho /\rho _{0}=0.1.$ The left panel of Figure 2 shows two iso-scaling
functions, $S(N)$, calculated with the predicted isotope yields for $3\leq
Z\leq 6$ for $\gamma =1$ and $\gamma =0$. In spite of a rather complex
interplay of expansion and fragment emission, the EES model predictions
still exhibit an approximate scaling. Some of the calculated deviations from
iso-scaling may, in fact, be due to the rather schematic treatment of the
Coulomb barrier penetration [22] and to the incomplete sequential decay
information used in the model. Values for $\alpha ,$ shown in the right
panel, are determined from the slopes of best fit of the lines to the
predicted iso-scaling functions.

For large values of $\gamma ,$ the asymmetry term, $C_{sym}\cdot (\rho /\rho
_{0})^{\gamma }$ decreases more rapidly with density and becomes negligible
as the residue expands. At low density where fragments are predominately
emitted [19], the difference in isotopic yields from the two reactions 1 and
2 become smaller, resulting in flatter scaling functions and smaller values
of $\alpha $. (In the extreme case of identical yields from the two systems, 
$S(N)$ becomes a horizontal line corresponding to $\alpha =0$.) The dot-dash
line in the right panel of Fig. 2 joining the solid points shows the EES
prediction that $\alpha $ decreases with increasing $\gamma $ values. The
multifragmentation data in Fig. 1 can be fairly well reproduced by $\gamma $ 
$\approx $0.6.

When the emission process ends at $\rho =0.1\rho _{0}$, a low density
residue (LDR) may remain. The final fate of this residue is not predicted by
the EES model. A potential ambiguity may result if this residue is large and
if its eventual disintegration produces a significant fraction of fragments
with $Z\geq 3$. Calculations using the Statistical Multifragmentation Model
[23] with the mass and energy of the final LDR predict $\alpha $-values that
increase with $\gamma $ (dashed line) --- opposite to the trend predicted
for emission during the expansion. When $\gamma $ decreases, the $N/Z$ of
the LDR from the two reactions, $^{124}Sn$ + $^{124}Sn$ and $^{112}Sn$ + $%
^{112}Sn$ become more similar, resulting in smaller $\alpha $-values -- a
trend also predicted from isospin dependent transport theory [24]. This
accounts for the behavior of the dashed line in the right panel of Figure 2.
If high energy fragments are emitted primarily from the expanding system,
and low-energy fragments come from the instantaneous disintegration of a
low-density residue, then the results in Fig. 2 suggest that the predicted
difference in the iso-scaling for low and high energy fragments should be
observable for $\gamma <0.8$.

In summary, we have observed a scaling between isotopic distributions which
allows a simple description of the dependence of such distributions on the
overall isospin of the measured systems in terms of three parameters, $%
\alpha ,\beta $ and $C.$ This scaling seems to apply to a broad range of
statistical fragment production mechanisms, including evaporation, strongly
damped binary collision, and multifragmentation. We have shown how this
systematics arises within models frequently applied to such processes. In
one such model, the EES model of multifragmentation, we find that the
iso-scaling parameters are sensitive to the density dependence of the
asymmetry term of the EOS.

This work was supported by the National Science Foundation under Grant Nos.
PHY-95-28844 and PHY-96-05140.

\newpage

\textbf{Figure Caption:}

\textbf{Figure 1: }The scaled isotopic ratio, $S(N)$ is plotted as a
function of the neutron number, N, using the best fit value of $\beta $
obtained from fitting isotopes with $Z\geq 3$. The data points plotted next
to the label ''multifragmentation'' in Figure 1 denote $S(N)$ extracted from
multifragmentation events in central $^{124}Sn+^{124}Sn$ and $%
^{112}Sn+^{112}Sn$ collisions[7] with $\alpha =0.37,\beta =-0.40$,. The
scaling behavior for evaporation process is illustrated by the reactions $%
^{4}He+^{116}Sn$ and $^{4}He+^{124}Sn$ [12] plotted next to the label
''evaporation'' with $\alpha =0.60,\beta =-0.82$. Systematics of the
strongly damped binary collisions is represented by the data of $^{16}O$
induced reactions on two targets $^{232}Th$ and $^{197}Au$ [11] plotted next
to the label ''deeply inelastic'' with $\alpha =0.74,\beta =-1.1$.

\textbf{Figure 2:} Theoretical EES model predictions for the scaling
functions $S(N)$ (left panel) and scaling parameter $\alpha $ (right panel)
of fragments emitted in $^{124}Sn$ + $^{124}Sn$ and $^{112}Sn$ + $^{112}Sn$
collisions. Same convention as Figure 1 applies to the various symbols used
in the left hand panel.

\newpage

\textbf{References:}

1. Bao-An Li et al., Phys. Rev. Lett. 78, 1644 (1997).

2. I. Bombaci, et al., Phys. Rep. 242, 165 (1994).

3. J.M. Lattimer and M. Prakash, Ap. J. (in press).

4. R. Laforest et al., Phys. Rev. C59 2567 (1999) and refs. therein.

5. R. Wada et. al., Phys. Rev. Lett. 58, 1829 (1987).

6. W.U. Schroeder and U.J. Huizenga, Treatise on heavy Ion Science, Ed. D.A.
Bromley (Plenum Press, 1984) and references therein.

7. H. Xu et al., Phys. Rev. Lett. 85, 716 (2000).

8. J. Randrup and S.E. Koonin, Nucl. Phys. A 356, 223 (1981).

9. S. Albergo et al., Nuovo Cimento A 89, 1 (1985).

10. G. J. Kunde et al., Phys. Lett. B416, 56 (1998). 

11. M. D'Agostino et al., Phys. Lett. B 371, 175 (1996); B.A. Li et al.,
Phys.Lett.B303, 225 (1993); A.S. Botvina et al., Nucl. Phys. A 584, 737
(1995).

12. V.V. Volkov, Phys. Rep. 44, 93, (1978).

13. J. Brzychczyk et al., Phys. Rev. C47, 1553 (1993).

14. C.K. Gelbke et. al., Phys. Rep. 42, 311 (1978).

15. S.R. Souza, W.P. Tan, R. Donangelo, C.K. Gelbke, W.G. Lynch, M.B. Tsang,
Phys. Rev. C62, 064607 (2000).

16. W. Friedman and W. Lynch, Phys. Rev. C 28, 950 (1983).

17. This approach is similar to the Weisskopf model of V. Weisskopf, Phys.
Rev, 52, 295 (1937).

18. V. E. Viola, K. Kwiatkowski, W. A. Friedman, Phys. Rev. C 59, 2660 (1999)

19. W.A. Friedman, Phys. Rev. Lett. 60, (1988) 2125; and Phys. Rev. C42, 667
(1990).

20. Note: $f^{*}(T/\epsilon _{f})/T$ remains roughly constant during
expansion.

21. A. Bohr \&\ B.\ R. Mottleson, ''Nuclear Structure, Vol II'', W.A.
Benjamin Inc., (1998).

22. J. T\~{o}ke, W. Gawlikowicz, and W. U. Schr\"{o}der, Phys. Rev. C 63,
024606 (2001).

23. A. Majumder and S. Das Gupta, Das Gupta, Phys. Rev. C 61, 034603 (2000),
Scott Pratt, Subal Das Gupta, Phys. Rev. C 62, 044603 (2000)

24. W. Tan et al, private communications and to be published.

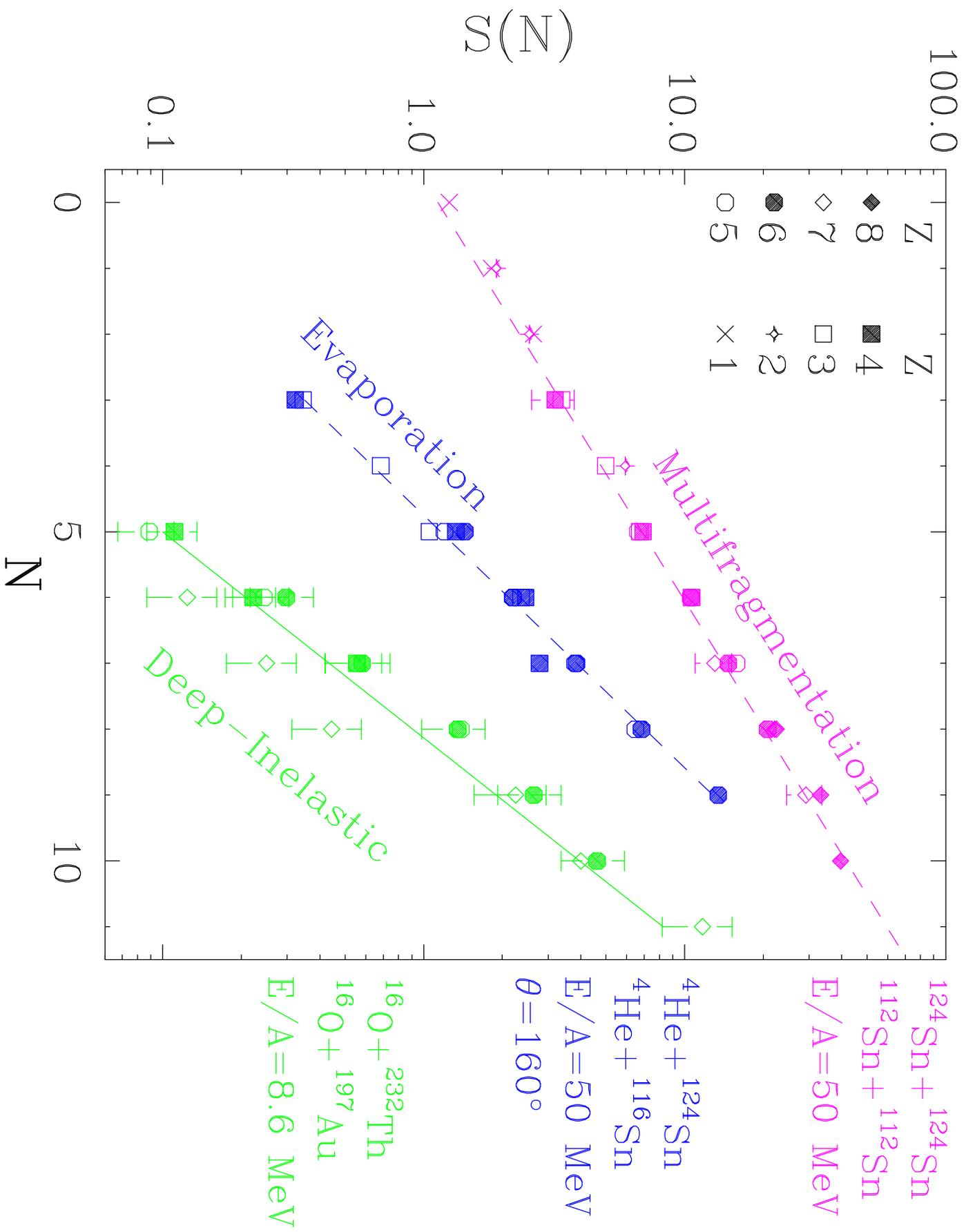
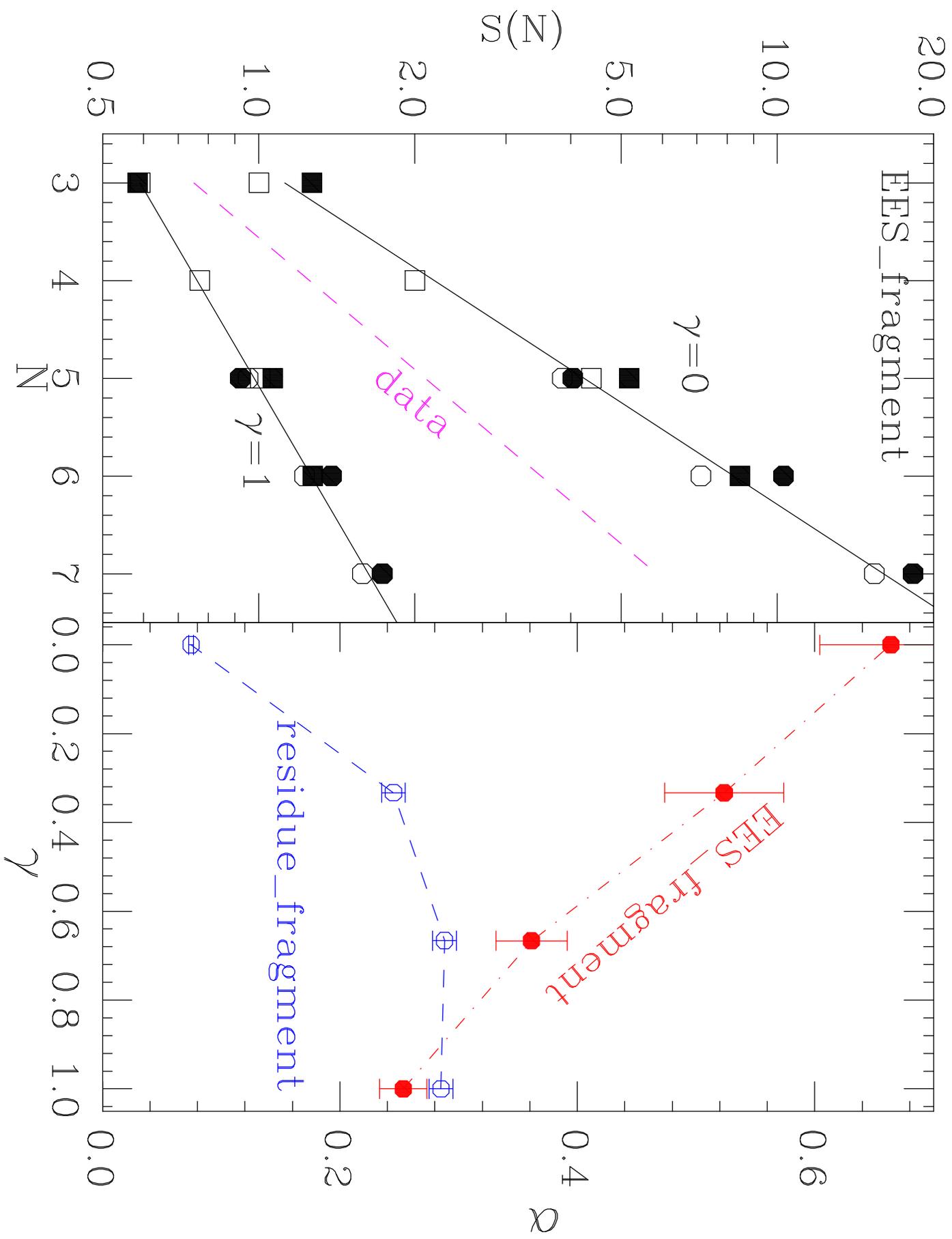
\end{document}